\documentclass[showkeys]{revtex4}

\usepackage{graphicx}
\usepackage{bm}
\usepackage{amssymb}
\usepackage{amsmath}
\usepackage{dcolumn}
\usepackage[latin1]{inputenc}
\usepackage[english]{babel}

\newcommand{\abs}[1]{\left \lvert #1 \right \rvert}
\newcommand{\bk}[1]{\left \langle #1 \right \rangle}

\begin{document}


\title{Compact equations for the envelope theory}

\author{Lorenzo \surname{Cimino}}
\email[E-mail: ]{lorenzo.cimino@umons.ac.be}
\thanks{ORCiD: 0000-0002-6286-0722}

\author{Claude \surname{Semay}}
\email[E-mail: ]{claude.semay@umons.ac.be}
\thanks{ORCiD: 0000-0001-6841-9850}

\affiliation{Service de Physique Nucl\'{e}aire et Subnucl\'{e}aire,
Universit\'{e} de Mons,
UMONS Research Institute for Complex Systems,
Place du Parc 20, 7000 Mons, Belgium}
\date{\today}

\begin{abstract}
\textbf{Abstract} The envelope theory is a method to easily obtain approximate, but reliable, solutions for some quantum many-body problems. Quite general Hamiltonians can be considered for systems composed of an arbitrary number of different particles in $D$ dimensions. In the case of identical particles, a compact set of 3 equations can be written to find the eigensolutions. This set provides also a nice interpretation and a starting point to improve the method. It is shown here that a similar set of 7 equations can be determined for a system containing an arbitrary number of two different particles.
\end{abstract}

\maketitle

\section{introduction}
\label{sec:intro}

The envelope theory (ET) \cite{hall80,hall83,hall04} is a technique to compute approximate eigenvalues and eigenvectors of $N$-body systems. This method, first developed for systems with identical particles, has been extended to treat non-standard kinematics in $D$ dimensions in \cite{sema13,sema19}, and it has been recently generalized for systems with different particles \cite{sema20}. The big advantage of this method is that the computation cost is independent of the number of particles. Quite general Hamiltonians can be considered, and the approximate eigenvalues are lower or upper bounds in favorable cases. The method relies on the existence of an exact solution for the $N$-body harmonic oscillator Hamiltonian \cite{hall79,cint01}. The accuracy of the method has been checked for various three-dimensional systems \cite{sema15a} and one-dimensional systems containing up to 100 bosons \cite{sema19}.

It is worth noting that the ET method has been rediscovered in 2008 under the name of the auxiliary field method, following an approach different from the one used by Hall \cite{hall80,hall83,hall04}. It has been recognized later that both methods are actually completely equivalent. This story is described in \cite{silv12}, where a lot of information is given about this approximation method. 

The ET has been used to obtain physical results about hadronic systems as in \cite{sema09}, and is especially useful when the number of particles can be arbitrary large as in the large-$N$ formulation of QCD \cite{buis11,buis12}. The method has allowed the study of a possible quasi Kepler's third law for quantum many-body systems \cite{sema21}. It can also be simply used to test accurate numerical calculations as in \cite{char15}. 
 
Let us consider the $N$-body Hamiltonian
\begin{equation}\label{trueH}
    H=\sum_{i=1}^N T_i(p_i) + \sum_{i<j=2}^N V_{ij}(r_{ij}),
\end{equation}
where $T_i$ is an arbitrary kinetic energy with some constraints \cite{sema18a} and $V_{ij}$ is a two-body central potential. We also define $p_i=\abs{\bm{p}_i}$ and $r_{ij}=\abs{\bm{r}_i-\bm{r}_j}$, where $\bm{r}_i$ and $\bm{p}_i$ are respectively the position and the momentum of the $i$th particle. It is assumed in the following that we are always working in the centre of mass (CM) frame, $\bm{P} = \sum_{i=1}^N \bm{p}_i=\bm{0}$, and with natural units ($\hbar=c=1$).

As explained in \cite{silv10,sema20}, in the framework of the ET, Hamiltonian (\ref{trueH}) is replaced by an auxiliary Hamiltonian (it is the origin of the other name of the method)
\begin{equation}\label{auxH}
    \tilde{H}(\{\alpha\})=\sum_{i=1}^N \left[\frac{\bm{p}_i^2}{2\mu_i}+T_i(G_i(\mu_i))-\dfrac{G_i^2(\mu_i)}{2\mu_i}\right] + \sum_{i<j=2}^N\left[ \rho_{ij}\bm{r}_{ij}^2+V_{ij}(J_{ij}(\rho_{ij}))-\rho_{ij}J_{ij}^2(\rho_{ij})\right],
\end{equation}
where $\{\alpha\} = \{\{\mu_i\},\{\rho_{ij}\}\}$ is a set of auxiliary parameters to determine later, and where the auxiliary functions $G_i$ and $J_{ij}$ are such that
\begin{equation}\label{deffunc}
    \begin{array}{cc}
        T'_i(G_i(x))-\dfrac{G_i(x)}{x}=0,\\[0.5cm]
        V'_{ij}(J_{ij}(x))-2xJ_{ij}(x)=0,
    \end{array}
\end{equation}
where $U'(x)=dU(x)/dx$. It is useful to write Hamiltonian (\ref{auxH}) in the form
\begin{equation}
    \tilde{H}(\{\alpha\}) = H_\text{ho}(\{\alpha\})+B(\{\alpha\}),
\end{equation}
where $H_\text{ho}$ is the harmonic oscillator part and $B$ is a function obtained by subtracting the harmonic oscillator contributions from (\ref{auxH}).
An eigenvalue of (\ref{auxH}) is given by
\begin{equation}\label{energy}
    \tilde{E}(\{\alpha\})=E_\text{ho}(\{\alpha\})+B(\{\alpha\}),
\end{equation}
where $E_\text{ho}$ is an eigenvalue of $H_\text{ho}$. A procedure in \cite{silv10,sema20} explains how to compute $E_\text{ho}$ but an example will be given below. An eigenvalue $\tilde{E}$ also depends on the set of parameters $\{\alpha\}=\{\{\mu_i\},\{\rho_{ij}\}\}$. The principle of the method is to search for the set of parameters $\{\alpha_0\}=\{\{\mu_{i0}\},\{\rho_{ij0}\}\}$ such that
\begin{equation}\label{mineq}
    \frac{\partial\tilde{E}}{\partial\mu_i}\biggr\rvert_{\{\alpha_0\}}=\frac{\partial\tilde{E}}{\partial\rho_{ij}}\biggr\rvert_{\{\alpha_0\}}=0 \hspace{5 mm} \forall\ i,j.
\end{equation}
Equations (\ref{mineq}) can be easily implemented and solutions $\{\alpha_0\}$ are easily found since we only need to find an extremum \cite{sema20}. After solving (\ref{mineq}), we obtain the desired approximate energy by substituting the set $\{\alpha_0\}$ back to (\ref{energy}),  $\tilde{E}(\{\alpha_0\})=\tilde{E}_0$.

In the case of identical particles, it has been showed \cite{sema13,sema19} that we can equivalently find the eigenvalue $\tilde{E}_0$ by using a set of three compact equations
    \begin{subequations}\label{compacteq}
             \begin{equation}\label{compacteq1}
                 \tilde{E}_0 = N\,T(p_0)+C^2_N\,V(\rho_0),
             \end{equation}
             \begin{equation}\label{compacteq3}
                 N\,T'(p_0)\,p_0=C^2_N\,V'(\rho_0)\,\rho_0,
             \end{equation}
             \begin{equation}\label{compacteq2}
                 Q(N) = \sqrt{C^2_N}\,p_0\,\rho_0,
             \end{equation}
    \end{subequations}
where $C^2_N=N(N-1)/2$ is the number of pairs, and where $p_0^2=\bk{\bm{p}_i^2}$ and $\rho_0^2 = \bk{\bm{r}_{ij}^2} \hspace{2mm} \forall\ i,j$. The mean values are taken with an eigenstate of the auxiliary Hamiltonian corresponding to the global quantum number $Q(N)$ for the set $\{\alpha_0\}$ insuring the constraints (\ref{mineq}). The eigenstate is also completely (anti)symmetric for the exchange between particles.
\begin{equation}
  Q(N)=
    \begin{cases}
      \sum\limits_{i=1}^{N-1} \left(2n_i+l_i+\frac{D}{2}\right) & \text{ if }D\geq2\\[15pt]
      \sum\limits_{i=1}^{N-1} \left(n_i+\frac{1}{2}\right) & \text{ if }D=1
    \end{cases},
\end{equation}
where the quantum numbers $\{n_i,l_i\}$ are associated with the internal Jacobi variables. Some values of $Q(N)$ for the bosonic and fermionic ground states are given in \cite{sema20,sema19}. In previous papers \cite{sema13,sema19}, the variable $r_0^2 = N^2 \bk{\left(\bm{r}_i-\bm{R}\right)^2}$, where $\bm{R}$ is the CM position, was used instead of $\rho_0$ because one-body and two-body potentials are treated together.

These equations are called compact because all the relevant variables appear in 3 equations giving the definition of the energy (\ref{compacteq1}), the equation of motion (\ref{compacteq3}) and the rule for the quantization (\ref{compacteq2}). Moreover, the uninteresting auxiliary parameters and functions are not present. Equations (\ref{compacteq}) can also be easily implemented and solved. There are good reasons to prefer the compact equations (\ref{compacteq}) over the ``extremization'' equations (\ref{mineq}). First, the quantities $p_0$ and $\rho_0$ give direct access to more interesting expectation values than $\{\alpha_0\}$. Secondly, these equations have a nice semiclassical interpretation as explained in \cite{sema13}. Thirdly, it is possible to improve the ET with the dominantly orbital state method starting from these equations \cite{sema15b}, which is the main motivation to write these equations. As the improvement obtained can be significant in some cases, it is worth generalizing it beyond systems of identical particles. In the following section, we will present the compact equations for a system composed of two different sets of $N_a$ and $N_b$ identical particles.

\section{$\bm{N_a + N_b}$ systems}
\label{sec:nanb}

Let us specify the auxiliary Hamiltonian (\ref{auxH}) for this system. The harmonic oscillator Hamiltonian for a system of $N_a$ particles of type $a$ and $N_b$ particles of type $b$ is given by
\begin{equation}\label{ho}
    H_\text{ho}=\sum_{i=1}^{N_a} \frac{\bm{p}_i^2}{2\mu_a}+\sum_{j=1}^{N_b} \frac{\bm{p}_j^2}{2\mu_b}+\sum_{i<i'=2}^{N_a}\rho_{aa}\bm{r}_{ii'}^2+\sum_{j<j'=2}^{N_b}\rho_{bb}\bm{r}_{jj'}^2+\sum_{i=1}^{N_a}\sum_{j=1}^{N_b}\rho_{ab}\bm{r}_{ij}^2.
\end{equation}
In the following, letters $i\text{ }(j)$ are reserved for particles of type $a\text{ }(b)$. As explained in \cite{sema20,hall79}, it is useful to write (\ref{ho}) in the form
    \begin{subequations}\label{ho2}
        \begin{equation}\label{ho20}
             H_\text{ho}=H_a+H_b+H_\text{CM} \hspace{1cm} \text{with}
        \end{equation}
        \begin{equation}\label{ho21}
            H_a = \sum_{i=1}^{N_a}{\frac{\bm{p}_i^2}{2\mu_a}}-\frac{\bm{P}_a^2}{2M_a}+\sum_{i<i'=2}^{N_a}\left(\rho_{aa}+\frac{N_b}{N_a}\rho_{ab}\right)\bm{r}_{ii'}^2,
        \end{equation}
        \begin{equation}\label{ho22}
            H_b = \sum_{j=1}^{N_b}{\frac{\bm{p}_j^2}{2\mu_b}}-\frac{\bm{P}_b^2}{2M_b}+\sum_{j<j'=2}^{N_b}\left(\rho_{bb}+\frac{N_a}{N_b}\rho_{ab}\right)\bm{r}_{jj'}^2,
        \end{equation}
        \begin{equation} \label{ho23}
            H_\text{CM} = \frac{\bm{p}^2}{2\mu}+N_aN_b\rho_{ab}\bm{r}^2,
        \end{equation}
    \end{subequations}
where $\bm{P}_\alpha$ and $M_\alpha=N_\alpha\,\mu_\alpha$ are the total momentum and mass for the set $\alpha = \{a,b\}$, $\mu=\frac{M_aM_b}{M_a+M_b}$ is a reduced mass, and $\bm{p}=\frac{M_b\bm{P}_a-M_a\bm{P}_b}{M_a+M_b}$ and $\bm{r}=\bm{R}_a-\bm{R}_b$ are the relative momentum and position between the CM of the two sets, respectively. The three parts of (\ref{ho20}) are entirely decoupled since (\ref{ho21}) and (\ref{ho22}) depends on the internal coordinates of their respective set, and (\ref{ho23}) on the relative coordinates between the two CM.  

Then, an eigenvalue $E_\text{ho}$ is easily obtained since (\ref{ho2}) is composed of three decoupled parts \cite{sema20}
\begin{equation}\label{enho}
        E_\text{ho}=Q(N_a)\sqrt{\frac{2}{\mu_a}(N_a\rho_{aa}+N_b\rho_{ab})}+Q(N_b)\sqrt{\frac{2}{\mu_b}(N_b\rho_{bb}+N_a\rho_{ab})}+Q(2)\sqrt{\frac{2}{\mu}N_aN_b\rho_{ab}}.
\end{equation}
To be complete, the expression of the function $B(\{\alpha\})$ is given by 
\begin{equation}\label{B}
    \begin{aligned}
    B & =  N_a\left[T_a(G_a(\mu_a))-\frac{G^2_a(\mu_a)}{2\mu_a}\right]+C^2_{N_a}\left[V_{aa}(J_{aa}(\rho_{aa}))-\rho_{aa}J^2_{aa}(\rho_{aa})\right]\\
    & +N_b\left[T_b(G_b(\mu_b))-\frac{G^2_b(\mu_b)}{2\mu_b}\right]+C^2_{N_b}\left[V_{bb}(J_{bb}(\rho_{bb}))-\rho_{bb}J^2_{bb}(\rho_{bb})\right] \\
    & +N_aN_b\left[V_{ab}(J_{ab}(\rho_{ab}))-\rho_{ab}J^2_{ab}(\rho_{ab})\right].
    \end{aligned}
\end{equation}
When combining (\ref{ho2}) and (\ref{B}), we can see that our auxiliary Hamiltonian (\ref{auxH}) is also composed of three distinct parts: one for the particles of type $a$, another for the particles of type $b$ and a last one for the relative motion between the two sets.

The compact equations can then be established in a similar way as done for identical particles \cite{silv12}. First, we apply the Hellmann-Feynman theorem \cite{hell} on Hamiltonian (\ref{auxH}) to evaluate extremization conditions (\ref{mineq}). By using definitions (\ref{deffunc}) we get the following results
\begin{equation}\label{hf}
    \begin{array}{lllll}
         & G_a^2(\mu_{a0})=p_a^2+\frac{P_0^2}{N_a^2} = {p^\prime_a}^2, \\[0.3cm]
         & G_b^2(\mu_{b0})=p_b^2+\frac{P_0^2}{N_b^2} = {p^\prime_b}^2, \\[0.3cm]
         & J^2_{aa}(\rho_{aa0})=r_{aa}^2, \\[0.3cm]
         & J^2_{bb}(\rho_{bb0})=r_{bb}^2, \\[0.3cm]
         & J^2_{ab}(\rho_{ab0})=\frac{N_a-1}{2N_a}r_{aa}^2+\frac{N_b-1}{2N_b}r_{bb}^2+R_0^2= {r^\prime_0}^2,
    \end{array}
\end{equation}
where we have defined the six physical parameters
\begin{equation}
    \begin{array}{lll}
         & p_a^2=\bk{\bm{p}_i^2-\frac{\bm{P}_a^2}{N_a^2}}\text{ and } p_b^2=\bk{\bm{p}_j^2-\frac{\bm{P}_b^2}{N_b^2}},\\[0.3cm]
         & r_{aa}^2= \bk{\bm{r}_{ii'}^2}\text{ and } r_{bb}^2= \bk{\bm{r}_{jj'}^2},\\[0.3cm]
         & P_0^2=\bk{\bm{p}^2}\text{ and } R_0^2=\bk{\bm{r}^2}.
    \end{array}
\end{equation}
The mean values are taken with an eigenstate of the auxiliary Hamiltonian corresponding to the quantum numbers $Q(N_a)$, $Q(N_b)$ and $Q(2)$ for the set $\{\alpha_0\}$ insuring the constraints (\ref{mineq}). The eigenstate is also completely (anti)symmetric for the exchange between the $N_a$ or the $N_b$ particles.

Then, by evaluating $\tilde{E}_0=\bk{\tilde{H}(\{\alpha_0\})}$ and using results (\ref{hf}), we obtain the following equation for the energy
\begin{equation}\label{eqen}
    \tilde{E}_0=N_aT_a\left(p'_a\right)+N_bT_b\left(p'_b\right)+C^2_{N_a}V_{aa}\left(r_{aa}\right)+C^2_{N_b}V_{bb}\left(r_{bb}\right)+N_aN_bV_{ab}\left(r_0'\right)
\end{equation}
It is interesting to look at the meaning of the linear combinations $p'_a$, $p'_b$ and $r'_0$ since they appear in (\ref{eqen}). As shown in \cite{sema20}, we can derive a similar equation for the energy by using the form (\ref{ho}), instead of (\ref{ho2}), of the harmonic oscillator. By comparing, one can identify ${p'_a}^2 = \bk{\bm{p}_i^2}$, ${p'_b}^2=\bk{\bm{p}_j^2}$ and ${r'_0}^2=\bk{\bm{r}_{ij}^2}$. 

In order to find these parameters, we need 6 additional equations. We can find three of them by applying the virial theorem separately on each of the three parts of the auxiliary Hamiltonian \cite{sema20}. One gets
\begin{subequations}\label{eqvirial}  
    \begin{equation}\label{eqvirial1}
        N_aT'_a(p'_a)\frac{p_a^2}{p'_a}=C^2_{N_a}V'_{aa}(r_{aa})r_{aa}+\frac{N_b}{N_a}C^2_{N_a}V'_{ab}(r'_0)\frac{r_{aa}^2}{r'_0},
    \end{equation}
    \begin{equation}\label{eqvirial2}
        N_bT'_b(p'_b)\frac{p_b^2}{p'_b}=C^2_{N_b}V'_{bb}(r_{bb})r_{bb}+\frac{N_a}{N_b}C^2_{N_b}V'_{ab}(r'_0)\frac{r_{bb}^2}{r'_0},
    \end{equation}
    \begin{equation}\label{eqvirial3}
       \frac{1}{N_a} T'_a(p'_a)\frac{P_0^2}{p'_a}+\frac{1}{N_b} T'_b(p'_b)\frac{P_0^2}{p'_b}=N_aN_bV'_{ab}(r'_0)\frac{R_0^2}{r'_0}.
    \end{equation}
\end{subequations}

Finally, we obtain three last equations by using the exact eigenvalue (\ref{enho}) of the harmonic oscillator and comparing it to $\bk{H_\text{ho}\{\alpha_0\}}$. Thanks to (\ref{ho2}), the comparison is done in a similar way as in \cite{silv12} and one gets 
\begin{subequations}\label{eqcomp} 
    \begin{equation}\label{eqcomp1}
        Q(N_a)=\sqrt{C^2_{N_a}}p_ar_{aa},
    \end{equation}
    \begin{equation}\label{eqcomp2}
       Q(N_b)=\sqrt{C^2_{N_b}}p_br_{bb},
    \end{equation}
    \begin{equation}\label{eqcomp3}
        Q(2)=P_0R_0.
    \end{equation}
\end{subequations}

Equations (\ref{eqvirial}) and (\ref{eqcomp}) form a set of six equations which, combined with (\ref{eqen}), form the ET compact equations for a system of $N_a+N_b$ particles, and allow us to compute the approximate eigenvalue $ \tilde{E}_0$. We have verified on several systems that these equations give the same results than those found with the extremization equations in \cite{sema20}. We can note that the three equations (\ref{eqvirial}) can be derived by minimizing (\ref{eqen}) with respect to $r_{aa}$, $r_{bb}$ and $R_0$, and using (\ref{eqcomp}). 

Equations (\ref{eqen})-(\ref{eqcomp}) are more complicated than equations (\ref{compacteq}). But, when comparing the two sets, it is possible to find an interpretation for equations (\ref{eqen})-(\ref{eqcomp}). Equation (\ref{eqen}) is obviously the energy computed in terms of the mean momenta and relative distances. Equations (\ref{eqvirial}) are the equations of motion determining these mean quantities, and equations (\ref{eqcomp}) are the semiclassical quantifications of the various orbital and radial motions. These equations make clear what are the relevant quantities appearing in a quantum system containing two different sets of identical particles. It is worth recalling that solutions obtained by the ET are full quantum ones with eigenfunctions associated \cite{silv10,sema20} and that observables can be computed \cite{sema15a}. 

As a first check for these equations, we need to recover the three equations (\ref{compacteq}) when considering all particles identical. In this case $T_a = T_b$ and $V_{aa}=V_{bb}=V_{ab}$, and we must impose the following symmetries
\begin{equation}\label{sym}
        \begin{array}{cc}
         & \bk{\bm{p}_i^2} = \bk{\bm{p}_j^2}, \hspace{5 mm} \forall\ i,j, \\[0.5cm]
         & \bk{\bm{r}_{ii'}^2} = \bk{\bm{r}_{jj'}^2} = \bk{\bm{r} _{ij}^2}, \hspace{5mm} \forall\ i,i',j,j'.
    \end{array}
\end{equation}
From the definitions of our 6 parameters, we conclude
\begin{equation}\label{sym2}
    \begin{array}{cc}
         & p'_a = p'_b = p_0, \\[0.5cm]
         & r_{aa} = r_{bb} = r'_0=\rho_0,
    \end{array}
\end{equation}
where $p_0$ and $\rho_0$ are defined as before in (\ref{compacteq}). Then, we easily see that equation (\ref{eqen}) reduces to (\ref{compacteq1}) with $N = N_a + N_b$. It is a matter of algebra to show that the sum of the three equations (\ref{eqvirial})
\begin{equation}
    N_aT'_a(p'_a)p'_a + N_bT'_b(p'_b)p'_b = C^2_{N_a}V'_{aa}(r_{aa})r_{aa} + C^2_{N_b}V'_{bb}(r_{bb})r_{bb} + N_aN_bV'_{ab}(r'_0)r'_0,
\end{equation}
reduces to (\ref{compacteq3}). 
When all the particles are identical, it is not relevant to separate the energy on several subsets. We notice that $Q(N_a)+Q(N_b)+Q(2) = Q(N_a+N_b)$. This is a hint that the sum of the three equations (\ref{eqcomp}) must reduce to (\ref{compacteq2}), but the proof is more subtle. Thanks to the symmetries  (\ref{sym}) and (\ref{sym2}), one can express $R_0$ in terms of $\rho_0$, and $p_a$, $p_b$ and $P_0$ in terms $p_0$. Then, simple calculations show that (\ref{eqcomp}) reduces to (\ref{compacteq2}). Finally, all equations (\ref{compacteq}) are recovered.
Note that (\ref{sym2}) also implies symmetries on the auxiliary parameters, $\mu_a = \mu_b$ and $\rho_{aa} = \rho_{bb} = \rho_{ab}$, which is also expected as explained in \cite{sema20}. 


As a second test, we have substituted the harmonic oscillator Hamiltonian (\ref{ho}) into our 7 equations. Then, it is a matter of algebra to find the exact solution (\ref{enho}). A third check is given in the following section. 

\section{$\bm{N_a=1}$ or/and $\bm{N_b=1}$}
\label{sec:na1}

The 7 equations (\ref{eqen}), (\ref{eqvirial}) and (\ref{eqcomp}) were computed for a system with $N_a+N_b$ particles. It is interesting to look at what happens when only one particle is present in a set. For example, let's look at the case $N_b=1$. Then, all the terms in $C^2_{N_b}$ and $Q(N_b)$ vanish. Equation (\ref{eqcomp2}) becomes trivial and (\ref{eqvirial2}) leads to $p_b = 0$. As $p_b=0$, we also have $p'_b = P_0$. At the end, we are left with a system of 5 equations
\begin{subequations}\label{eqn+1}
    \begin{equation}
        \tilde{E}_0=N_aT_a\left(p'_a\right)+T_b\left(P_0\right)+C^2_{N_a}V_{aa}\left(r_{aa}\right)+N_aV_{ab}\left(r_0'\right),
    \end{equation}
    \begin{equation}\label{eqn+12}
        N_aT'_a(p'_a)\frac{p_a^2}{p'_a}=C^2_{N_a}V'_{aa}(r_{aa})r_{aa}+\frac{N_a-1}{2}V'_{ab}(r'_0)\frac{r_{aa}^2}{r'_0},
    \end{equation}
    \begin{equation}
       \frac{1}{N_a} T'_a(p'_a)\frac{P_0^2}{p'_a}+T'_b(P_0)P_0=N_aV'_{ab}(r'_0)\frac{R_0^2}{r'_0},
    \end{equation}
    \begin{equation}\label{eqn+14}
        Q(N_a)=\sqrt{C^2_{N_a}}p_ar_{aa},
    \end{equation}
    \begin{equation}
        Q(2)=P_0R_0,
    \end{equation}
\end{subequations}
where our four parameters are now defined as $p_a^2=\bk{\bm{p}_i^2-\frac{\bm{P}_a^2}{N_a^2}}$, $P_0^2=\bk{\left(\frac{\mu_b\bm{P}_a-M_a\bm{p}_b}{M_a + \mu_b}\right)^2}$, $\bm{r}_{aa}^2= \bk{\bm{r}_{ii'}^2}$ and $R_0^2=\bk{\left(\bm{R}_a-\bm{r}_b\right)^2}$. We also have ${p'_a}^2=p_a^2+\frac{P_0^2}{N_a^2}$ and ${r'_0}^2=\frac{N_a-1}{2N_a}r_{aa}^2+R_0^2$. The five equations (\ref{eqn+1}) can also be found from scratch with the above explained  procedure.

Another special case is when $N_a=N_b=1$, that is we have a two-body system. We then have similar simplifications as in the previous case and we obtain the equations of the envelope theory at $N=2$, which are a generalization of the results obtained in  \cite{sema13,sema12}
\begin{subequations}
    \begin{equation}
        \tilde{E}_0=T_a(P_0)+T_b(P_0) + V_{ab}(R_0),
    \end{equation}
    \begin{equation}
        T'_a(P_0)P_0 + T'_b(P_0)P_0 = V'_{ab}(R_0)R_0,
    \end{equation}
    \begin{equation}
        Q(2)=P_0R_0,
    \end{equation}
\end{subequations}
where $P_0^2=\bk{\left(\frac{\mu_b\bm{p}_a-\mu_a\bm{p}_b}{\mu_a + \mu_b}\right)^2}$ and $R_0^2=\bk{\left(\bm{r}_a-\bm{r}_b\right)^2}$. The fact that the correct limits are obtained for $N_a=1$ or $N_b=1$ is also a test of coherence for the set~(\ref{eqen})-(\ref{eqvirial})-(\ref{eqcomp}). 

\section{concluding remarks}
\label{sec:conclu}

We were able to build the 7 compact equations of the envelope theory for a system of $N_a+N_b$ particles. These equations reduce to the 3 usual ones when considering identical particles. We also presented the special cases when $N_b =1$ or/and $N_a=1$. Starting from these equations, it is possible to improve the envelope theory using a similar procedure than the one used in \cite{sema15b}. This is performed for a system of $N_a+1$ particles in \cite{chev21}. With these 7 equations, it is possible to open new domains of applicability of the envelope theory, especially in hadronic physics where the method is proven to be useful as mentioned in the introduction. But the method can also be used to estimate the binding energies of other systems such as nuclei or clusters of cold atoms for which ab-initio calculations are already available, as for instance in \cite{gatt13}. In particular, accurate calculations have been performed for large helium clusters \cite{gatt11,kiev20}. For such systems, it is necessary to take into account a three-body forces that can be handled by the envelope theory \cite{sema18b}. 

\begin{acknowledgments}
L.C. would thank the Fonds de la Recherche Scientifique - FNRS for the financial support. This work was also supported under Grant Number 4.45.10.08.
\end{acknowledgments}


\begin{thebibliography}{99} 

\section*{References}

\bibitem{hall80} R.L. Hall, Energy trajectories for the $N$-boson problem by the method of potential envelopes. Phys. Rev. D \textbf{22}, 2062 (1980)
\bibitem{hall83} R.L. Hall, A geometrical theory of energy trajectories in quantum mechanics. J. Math. Phys. \textbf{24}, 324 (1983) 
\bibitem{hall04} R.L. Hall, W. Lucha, F.F. Sch\"oberl, Relativistic $N$-boson systems bound by pair potentials $V(r_{ij}) = g(r^2_{ij})$. J. Math. Phys. \textbf{45}, 3086 (2004)
\bibitem{sema13} C. Semay, C. Roland, Approximate solutions for $N$-body Hamiltonians with identical particles in $D$ dimensions. Res. Phys. \textbf{3}, 231 (2013)
\bibitem{sema19} C. Semay, L. Cimino, Tests of the Envelope Theory in One Dimension. Few-Body Syst. \textbf{60}, 64 (2019)
\bibitem{sema20} C. Semay, L. Cimino, C. Willemyns, Envelope theory for systems with different particles. Few-Body Syst. \textbf{61}, 19 (2020)
\bibitem{hall79} R.L. Hall, B. Schwesinger, The complete exact solution to the translation-invariant $N$-body harmonic oscillator problem. J. Math. Phys. \textbf{20}, 2481 (1979)
\bibitem{cint01} C.T. Willemyns, C. Semay, Some specific solutions to the translation-invariant $N$-body harmonic oscillator Hamiltonian. J. Phys. Commun. \textbf{5}, 115002 (2021)
\bibitem{sema15a} C. Semay, Numerical Tests of the Envelope Theory for Few-Boson Systems. Few-Body Syst. \textbf{56}, 149 (2015)
\bibitem{silv12} B. Silvestre-Brac, C. Semay, F. Buisseret, The Auxiliary Field Method in Quantum Mechanics. J. Phys. Math. \textbf{4}, P120601 (2012)
\bibitem{sema09} C. Semay, F. Buisseret, B. Silvestre-Brac, Towers of hybrid mesons. Phys. Rev. D \textbf{79}, 094020 (2009)
\bibitem{buis11} F. Buisseret, C. Semay, Light baryon masses in different large-$N_c$ limits. Phys. Rev. D \textbf{82}, 056008 (2010)
\bibitem{buis12} F. Buisseret, N. Matagne, C. Semay, Spin contribution to light baryons in different large-$N$ limits. Phys. Rev. D \textbf{85}, 036010 (2012)
\bibitem{sema21} C. Semay, C.T. Willemyns, Quasi Kepler's third law for quantum many-body systems. Eur. Phys. J. Plus \textbf{136}, 342 (2021)
\bibitem{char15} Y. Chargui, A. Dhahbi, A. Trabelsi, Exact analytical treatment of the asymmetrical spinless Salpeter equation with a Coulomb-type potential. Phys. Scr. \textbf{90}, 015201 (2015)
\bibitem{sema18a} C. Semay, Three theorems of quantum mechanics and their classical counterparts. Eur. J. Phys. \textbf{39}, 055401 (2018)
\bibitem{silv10} B. Silvestre-Brac, C. Semay, F. Buisseret, F. Brau, The quantum $\mathcal{N}$-body problem and the auxiliary field method. J. Math. Phys. \textbf{51}, 032104 (2010)
\bibitem{sema15b} C. Semay, Improvement of the envelope theory with the dominantly orbital state method. Eur. Phys. J. Plus \textbf{130}, 156 (2015)
\bibitem{hell} H. Hellmann, Ein kombiniertes Naherungsverfahren zur Energieberechnung im Vielelektronenproblem (German). Acta
Physicochimica U.R.S.S. \textbf{1}, 913 (1935)
\bibitem{sema12} C. Semay, An upper bound for asymmetrical spinless Salpeter equations. Phys. Lett. A \textbf{376}, 2217 (2012)
\bibitem{chev21} C. Chevalier, C.T. Willemyns, L. Cimino, C. Semay, Improvement of the Envelope Theory for Systems with Different Particles. arXiv:2111.14744
\bibitem{gatt13} M. Gattobigio, A. Kievsky, M. Viviani, Six-Bodies Calculations Using the Hyperspherical Harmonics Method, Few-Body Syst. \textbf{54}, 657 (2013)
\bibitem{gatt11} M. Gattobigio, A. Kievsky, M. Viviani, Spectra of helium clusters with up to six atoms using soft-core potentials, Phys. Rev. A \textbf{84}, 052503 (2011)
\bibitem{kiev20} A. Kievsky, A. Polls, B. Juli\'{a}-D\'{i}az, N.K. Timofeyuk, M. Gattobigio, Few bosons to many bosons inside the unitary window: A transition between universal and nonuniversal behavior, Phys. Rev. A \textbf{102}, 063320 (2020) 
\bibitem{sema18b} C. Semay, G. Sicorello, Many-Body Forces with the Envelope Theory. Few-Body Syst. \textbf{59}, 119 (2018)

\end{thebibliography}
\end{document}